\documentclass[]{aa}
\usepackage[nolist]{acronym}

\usepackage{graphicx}
\usepackage{txfonts}
\usepackage{natbib,twoopt}
\bibpunct{(}{)}{;}{a}{}{,}
\usepackage{hyperref}
\usepackage{nameref}
\usepackage{cleveref}
\crefname{figure}{Fig.}{Figs.}
\Crefname{figure}{Fig.}{Figs.}
\usepackage{nccmath}
\usepackage{subcaption}
\usepackage{siunitx}
\usepackage{booktabs}
\usepackage{multirow}
\usepackage{threeparttable}
\usepackage{tabularx}
\usepackage{array}
\usepackage{bm} 
\usepackage{float}
\usepackage[hyphenbreaks]{breakurl}
\usepackage{scalerel}
\usepackage{tikz}

\definecolor{cobalt}{rgb}{0.06, 0.2, 0.65}
\hypersetup{
  colorlinks,
  citecolor=cobalt,
  linkcolor=[rgb]{0.8, 0.2, 1.0},
  urlcolor=cobalt,
}

\makeatletter
\renewcommand*\aa@pageof{, page \thepage{} of \pageref*{LastPage}}

\newcommand{\logg}{\ensuremath{\log g}}

\def\kms{\ifmmode{\rm km\th s^{-1}}\else km\th s$^{-1}$\fi}
\def\th{\thinspace}

\newcommand{\twelveCO}{\textsuperscript{12}CO}
\newcommand{\thirteenCO}{\textsuperscript{13}CO}
\newcommand{\thirteenC}{\textsuperscript{13}C}

\newcommand{\Htwo}{H$_2$}
\newcommand{\pRT}{\texttt{petitRADTRANS}}

\newcommand{\micron}{$\mu$m}
\newcommand{\vsini}{$v\sin{i}$\ }
\newcommand{\Teff}{$T_{\text{eff}}$}

\newcommand{\Cratio}{\textsuperscript{12}C/\textsuperscript{13}C}

\newcommand{\water}{H$_2$O}
\newcommand{\methane}{CH$_4$}

\newcommand{\CRIRES}{CRIRES$^+$}

\newcommand{\fsed}{$f_{\rm sed}$}

\newcommand{\betapicb}{$\beta$\,Pic\,b}
\newcommand{\betapicc}{$\beta$\,Pic\,c}

\newcommandtwoopt{\citeads}[3][][]{\href{http://adsabs.harvard.edu/abs/#3}%
{\def\hyper@linkstart##1##2{}%
    \let\hyper@linkend\@empty\citealp[#1][#2]{#3}}}
\newcommandtwoopt{\citepads}[3][][]{\href{http://adsabs.harvard.edu/abs/#3}%
{\def\hyper@linkstart##1##2{}%
    \let\hyper@linkend\@empty\citep[#1][#2]{#3}}}

\makeatother

\begin{document} 

\title{The carbon isotope ratio of \betapicb\ with high-resolution spectroscopy}

\titlerunning{The carbon isotope ratio of \betapicb}
\author{
D. Gonz\'alez Picos\inst{1}
\and I.A.G. Snellen\inst{1}
\and R. Landman\inst{1}
\and S. de Regt\inst{1}
\and N. Grasser\inst{1}
\and J. L. Birkby\inst{3}
\and T. Stolker\inst{1}
\and I. Koutalios\inst{1}
\and M.A. Kenworthy\inst{1}
}

\institute{
Leiden Observatory, Leiden University, P.O. Box 9513, 2300 RA Leiden, The Netherlands\\
\email{picos@strw.leidenuniv.nl}
\and
Astrophysics, University of Oxford, Denys Wilkinson Building, Keble Road, Oxford, OX1 3RH, UK
}

   \date{Received YYYY-MM-DD; accepted YYYY-MM-DD}

\abstract
{Isotopic ratios trace the formation and evolution of planets and link their atmospheres to the chemistry of their natal protoplanetary discs. We measure \Cratio$\,=58^{+18}_{-15}$ in the atmosphere of the young super-Jupiter \betapicb\ from 11 nights of \CRIRES\ K-band spectroscopy ($\mathcal{R}\approx100{,}000$) at the Very Large Telescope (VLT). We detect both \twelveCO\ and \thirteenCO\ and constrain \Cratio\ with a Bayesian retrieval jointly fitted with near-infrared photometry. The inferred \Cratio\ is consistent with the present-day interstellar medium (ISM), is below the solar value, and is comparable to measurements in other young super-Jupiters. We also retrieve \Teff$\,=1629^{+30}_{-28}$\,K, near-solar to mildly super-solar metallicity ([M/H]$\,=0.20^{+0.16}_{-0.12}$), a solar-like carbon-to-oxygen ratio (C/O$\,=0.52\pm0.03$), and tentative evidence for thick clouds. We analyse each night independently and combine the results of the six epochs with the highest signal-to-noise ratio (S/N), propagating night-to-night scatter into the final uncertainties. This provides an isotopic benchmark for a directly-imaged planet interior to the CO snow line.}
  
\maketitle

\section{Introduction}\label{sec:introduction}

Beyond elemental abundances, isotopic ratios trace the chemical environment of the protoplanetary disc and the relative roles of gas and solids during accretion. A familiar Solar System example is the deuterium-to-hydrogen ratio (D/H): Jupiter and Saturn have D/H consistent with the protosolar nebula, whereas Uranus and Neptune are enriched, which is commonly attributed to accretion of D-rich ices beyond the water snow line \citep{feuchtgruberRatioAtmospheresUranus2013}. Carbon isotopes provide an analogous diagnostic for exoplanets. In discs, fractionation can shift \Cratio\ away from the local interstellar medium (ISM) value of approximately 68 \citep{milam12C13CIsotope2005}, for instance through isotopic ion-exchange reactions \citep{langer12C13CIsotope1993} and isotope-selective photodissociation of CO by ultraviolet radiation \citep{visserPhotodissociationChemistryCO2009}. These processes can enrich \thirteenC\ in ices relative to the gas, so a planet's atmospheric \Cratio\ depends on where it formed and on the balance between gas and solid carbon accretion \citep{zhang13COrichAtmosphereYoung2021,berginCarbonIsotopicRatio2024}.

High-resolution ($\mathcal{R}\gtrsim30{,}000$) near-infrared spectroscopy can measure \Cratio\ in substellar atmospheres. The CO overtone bands in the K band around 2.3\,\micron\ include resolved lines from both \twelveCO\ and \thirteenCO, allowing retrievals to disentangle their relative abundances \citep{molliereDetectingIsotopologuesExoplanet2019,zhang13COrichAtmosphereYoung2021}. The first exoplanet measurement targeted the young super-Jupiter YSES 1 b and initially reported \Cratio$\,=31^{+17}_{-10}$ \citep{zhang13COrichAtmosphereYoung2021}; follow-up work revised this to \Cratio$\,=88\pm13$ \citep{zhangESOSupJupSurvey2024}, highlighting the need for high resolution and adequate signal-to-noise ratio (S/N) when constraining isotopic ratios.

Improved observations and retrieval methods have made \Cratio\ measurements feasible for an increasing range of substellar objects and cool stars \citep{crossfieldUnusualIsotopicAbundances2019, zhang12CO132021, gandhiJWSTMeasurements13C2023, xuanAreThesePlanets2024,grasserESOSupJupSurvey2025, gonzalezpicosChemicalEvolutionImprints2025a}. Access to the fundamental CO band at 4.6\,\micron\ with the JWST has enabled precise measurements of \Cratio\ in a number of directly-imaged planets and brown dwarfs, including three of the four HR\,8799 planets \citep{ruffioJupiterlikeUniformMetal2026}, providing a direct view of how \Cratio\ varies with orbital radius within a single system. Several isotopic studies of young and field brown dwarfs have reported \Cratio\ values ranging from ISM-like to solar and above \citep{xuanValidationElementalIsotopic2024, regtESOSupJupSurvey2024, grasserESOSupJupSurvey2025, regtESOSupJupSurvey2026}. Observations of M dwarfs have also provided valuable context for the interpretation of isotopic ratios, suggesting that the wide spread of \Cratio\ in nearby M dwarfs is due to long-term Galactic chemical evolution \citep{gonzalezpicosChemicalEvolutionImprints2025a}, which can produce high \Cratio\ at old ages (\Cratio$\,\geq 200$ at $\approx 10$\,Gyr) and low \Cratio$\,\approx 60$ for young, metal-rich objects in the solar neighbourhood \citep{romanoEvolutionCNOElements2022}.

\begin{figure*}[!ht]
    \centering
    \includegraphics[width=\textwidth]{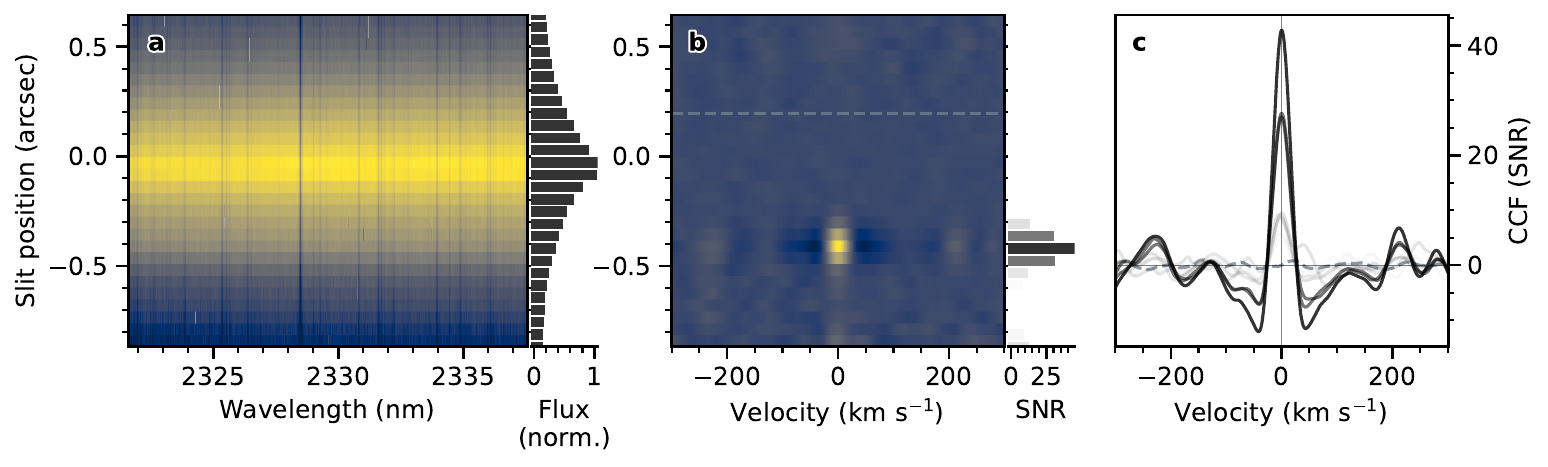}
    \caption{Detection of \betapicb. Panel a. Calibrated detector image of one echelle order with central wavelength 2330 nm. The scattered starlight is visible as a bright stripe close to the centre of the slit. Panel b. Cross-correlation of the data and a planet template after starlight subtraction across the slit. Panel c. Cross-correlation function at locations with significant planet signal. Line transparency follows the S/N per row in Panel b. The dashed line shows the cross-correlation on the opposite side of the slit, indicating the typical noise level.}
    \label{fig:fig1}
\end{figure*}

\section{Observations}
We observed \betapicb\ with \CRIRES\ at the VLT over 11 nights between October 2024 and May 2025, integrating one hour per night at $\mathcal{R}\approx100{,}000$ in the K band. The data were taken under programme 114.27DX.001 (PI: M. Kenworthy; \citealt{kenworthyBetaPic2026}) with a fixed instrument setup across all epochs. Unlike previous \CRIRES\ observations \citep{landmanPictorisEyesUpgraded2024}, we placed the star outside the slit (45\,$\deg$ slit angle) to improve the star-to-planet flux ratio and limit detector noise, allowing 120\,s exposures instead of 10\,s. Under favourable conditions (low airmass and seeing), the planet-to-star contrast is up to $\sim$4 times higher than in that analysis, which we attribute to the reduced detector noise and improved suppression of scattered starlight. Combined with our novel spectral extraction (see below), we obtain S/N per pixel at native resolution between 2 and 9 across nights, well above the $\lesssim1$ reported previously \citep{landmanPictorisEyesUpgraded2024}, making this the highest-S/N high-resolution spectrum of \betapicb\ to date.

We reduced the data with the \texttt{excalibuhr} pipeline \citep{zhangESOSupJupSurvey2024} (see also \citealt{regtESOSupJupSurvey2024,gonzalezpicosESOSupJupSurvey2025}). The reduction included dark and flat corrections, spectral-order tracing, lamp-based dispersion calibration applied to the science frames, nod-subtraction sky removal, and stacking of the frames acquired at each nod position. We then constructed a one-dimensional spectrum at the planet location using a novel spectral-extraction scheme tailored to faint sources on a strong slit background, such as close-in companions. Spatial rows with significant planetary signal were co-added with weights derived from the cross-correlation strength along the slit (\Cref{fig:fig1}). We calculated the cross-correlation map by cross-correlating a planetary template (\Teff=1700\,K, \logg=4.0, [M/H]=0.0; \citealt{morleySonoraSubstellarAtmosphere2024}) with the data at each row position after subtracting the fitted stellar model. The stellar model was jointly fitted with the planetary spectrum at every slit position using the linear model outlined in Sect.~\ref{sec:bayesian_retrieval}.

\section{Methods}

High-resolution spectroscopy resolves individual molecular lines and is sensitive to the pressure--temperature (P--T) structure. K-band observations probe \water\ and CO well but constrain clouds poorly: the narrow bandpass and lack of continuum information make it difficult to separate cloud opacity from metallicity and P--T effects. For \betapicb, this yields a strong metallicity--cloud degeneracy in spectroscopic analyses (e.g. \citealt{landmanPictorisEyesUpgraded2024}). We therefore fit the \CRIRES\ spectra jointly with archival 1.0--5.0\,\micron\ broadband photometry \citep{morzinskiMAGELLANADAPTIVEOPTICS2015}. The Y, J, and H bands respond to \Teff\ and cloud opacity, whereas the L and M bands are less cloud-sensitive and anchor the bolometric luminosity. Combining the line-resolved spectrum with the photometric spectral energy distribution (SED) tightens constraints on the atmospheric structure and clouds compared with \CRIRES\ alone, as shown for other directly-imaged planets whose continuum is lost in reduction (e.g. \citealt{ruffioJupiterlikeUniformMetal2026}). A limitation is that the photometry is single-epoch and not contemporaneous with the spectroscopy, which could bias the retrieval if the atmosphere is highly variable. \betapicb\ is nevertheless expected to be relatively stable ($\lesssim5\%$) as it is an early-L object, based on L-dwarf variability studies \citep{vosSpitzerVariabilityProperties2020,vosLetGreatWorld2022}, and cloud properties are not expected to significantly affect carbon isotope measurements \citep{xuanAreThesePlanets2024,regtESOSupJupSurvey2026}.

\begin{figure*}[ht]
    \centering
    \includegraphics[width=\linewidth]{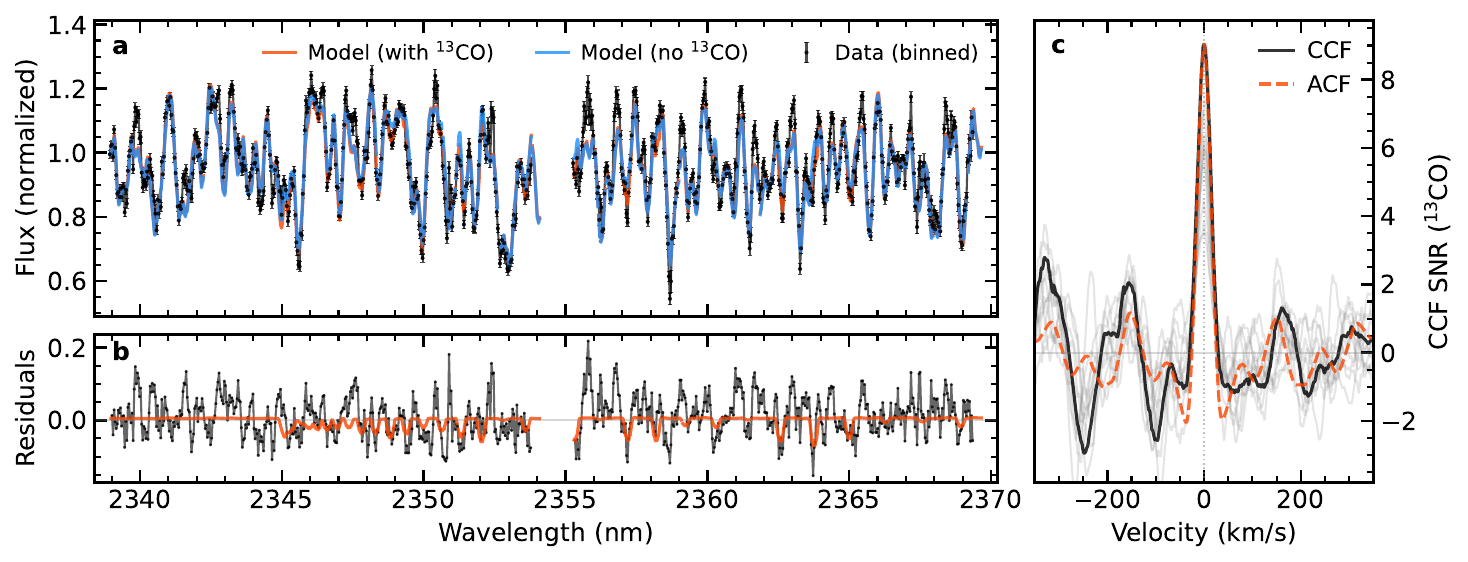}
    \caption{Detection of \thirteenCO\ in the atmosphere of \betapicb. Panel a. Combined observed spectrum from the six nights with S/N\,$>$\,5 in the \thirteenCO\ overtone bands, shifted to the planet rest-frame and binned to 10\,\kms\ for display. The two models are averages of the individual best-fit models with and without \thirteenCO. Panel b. Residuals between the observed spectrum and the model without \thirteenCO, with a \thirteenCO-only model overplotted. Panel c. Cross-correlation of the residuals, averaged over the full wavelength range and the six nights, with a \thirteenCO\ template. The dashed line shows the template auto-correlation function, and the faint grey lines show the individual-night cross-correlations.}
    \label{fig:fig_spectrum_13CO_chips10-11_A}
\end{figure*}

\begin{figure*}[ht]
    \centering
    \includegraphics[width=\linewidth]{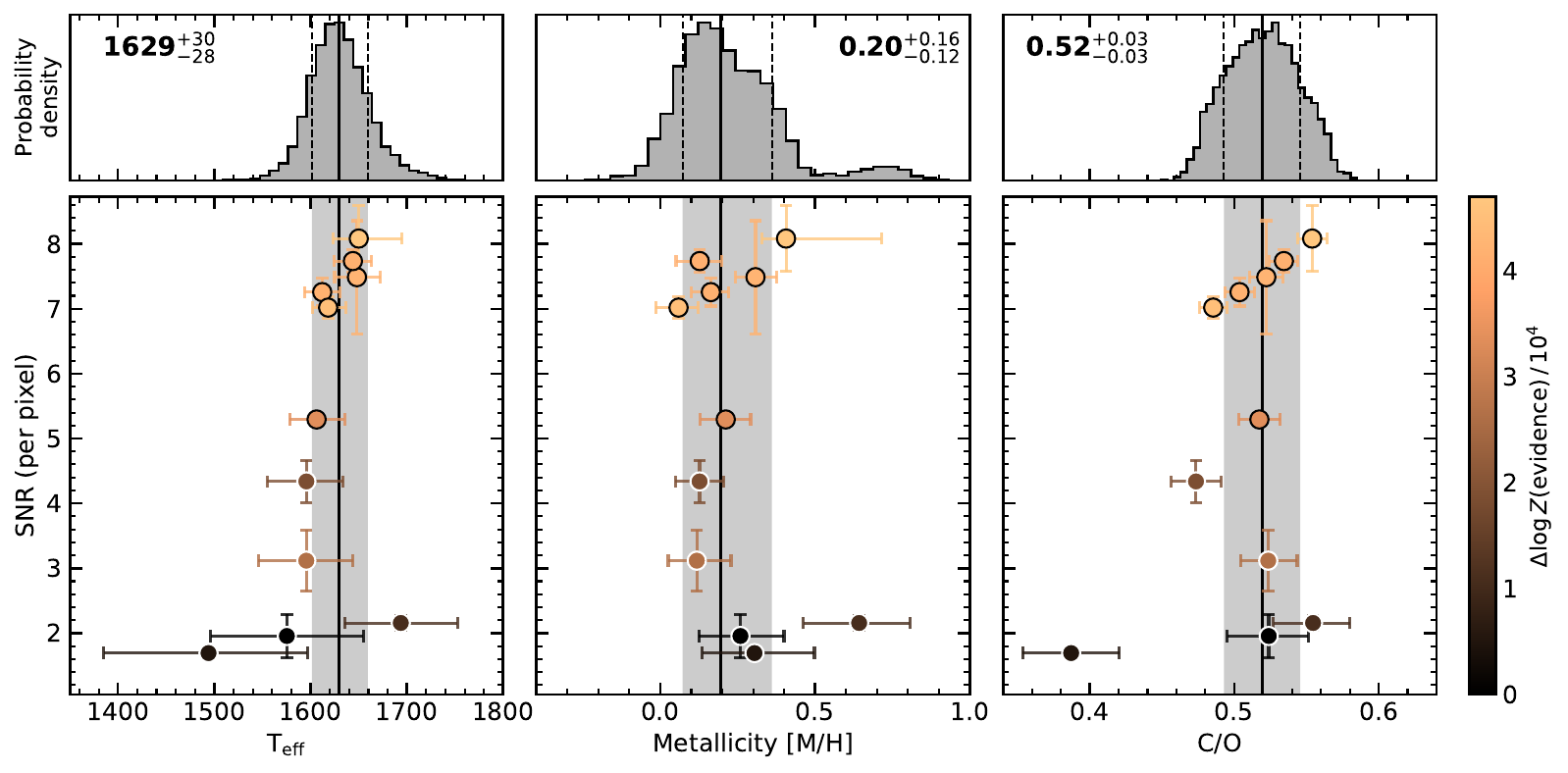}
    \caption{Retrieved atmospheric parameters. We show each individual night and the combined posterior of the six nights with S/N\,$>$\,5. The \Teff\ converges to a value consistent with the spectral type of \betapicb. The metallicity shows scatter and a bimodal tendency, which we attribute to the cloud--metallicity degeneracy. The C/O ratio is less affected by this degeneracy but still shows substantial epoch-to-epoch scatter.}
    \label{fig:fig_bma_teff_metallicity_c_to_o}
\end{figure*}

\begin{figure}[ht]
    \centering
    \includegraphics[width=\linewidth]{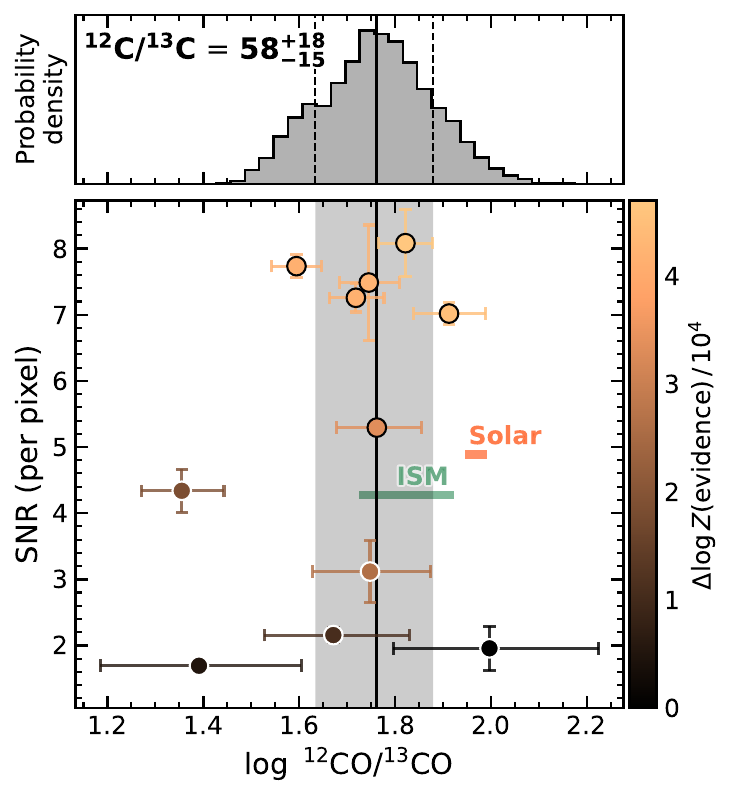}
    \caption{Measurement of \Cratio\ in the atmosphere of \betapicb. Top: combined posterior from the six nights with S/N per pixel\,$>$\,5. Bottom: per-night posteriors. The S/N uncertainties are estimated from the difference between the median S/N in the two nod positions. Colours encode the nested-sampling evidence of each run relative to the lowest-evidence case; higher values indicate stronger support.}
    \label{fig:fig_bma_log_12CO_13CO}
\end{figure}

\subsection{Forward model}

The observed K-band spectrum contains contributions from the planet, the partially scattered stellar point spread function, and telluric absorption by Earth's atmosphere. At the 0.48\arcsec\ projected separation of \betapicb, the raw starlight-to-planet flux ratio is approximately 100, so the star overwhelms the planet signal. Rather than subtracting it empirically, we included all three components simultaneously in a forward model, following an approach similar to that of \citet{gonzalezpicosESOSupJupSurvey2025} and inspired by \citet{landmanPictorisEyesUpgraded2024}.

We generated planetary spectra with \pRT, using P--T profiles drawn\footnote{\href{https://docs.scipy.org/doc/scipy/reference/generated/scipy.interpolate.RegularGridInterpolator.html}{scipy.interpolate.RegularGridInterpolator}} from the Sonora Diamondback grid of model atmospheres \citep{morleySonoraSubstellarAtmosphere2024}. This choice couples the thermal structure to the cloud properties and chemical equilibrium in a physically motivated way. The grid spans effective temperature \Teff, surface gravity \logg, metallicity [M/H], the sedimentation parameter $f_{\rm sed}$, which controls how efficiently cloud particles rain out of the atmosphere, and the eddy diffusion coefficient $K_{zz}$, which regulates vertical mixing of the cloud material. We obtained the atmospheric composition assuming chemical equilibrium, including condensation and rainout \citep{kitzmannFastchemCondEquilibrium2023}. Molecular line opacities for \twelveCO, \thirteenCO, \water, \methane, and HF were included alongside collision-induced absorption from \Htwo--\Htwo\ and \Htwo--He pairs (see \citealt{gonzalezpicosESOSupJupSurvey2025} for linelists and references). Condensate opacity was included using the parametrisation of \citealt{molliereRetrievingScatteringClouds2020a}, with the cloud particle size set by the eddy diffusion coefficient \citep{ackermanPrecipitatingCondensationClouds2001} and a log-normal size distribution set by the standard deviation $\sigma_{\rm g}$. We considered Fe and Mg$_2$SiO$_4$ condensates as possible cloud-forming species \citep{gaoAerosolCompositionHot2020,vosPatchyForsteriteClouds2023}. We included three independent free parameters for each species: cloud-base pressure $P_{\rm base, s}$, cloud-base mass fraction $X_{\rm base, s}$, and sedimentation parameter $f_{\rm sed, s}$. The model spectrum was Doppler-shifted, rotationally broadened by direct integration over the planetary surface \citep{carvalhoSimpleCodeRotational2023} using the projected rotational velocity \vsini, and convolved with a Gaussian kernel matching the instrumental resolution.

We extracted the telluric transmission empirically from the data. Because $\beta$\,Pictoris is an early-type star \citep{crifoPictorisRevisitedHipparcos1997} with a spectrum that closely resembles a blackbody in the K band, dividing the observed stellar spectrum by a Planck function at the stellar effective temperature yields a high-quality telluric absorption spectrum for each night (similar to \citealt{landmanPictorisEyesUpgraded2024}). We accounted for residual wavelength-dependent variation in the stellar contamination across the slit with a linear-spline scaling function fitted simultaneously with the stellar model, following \citet{gonzalezpicosESOSupJupSurvey2025}. This differs from the principal component analysis (PCA) approach used by \citet{landmanPictorisEyesUpgraded2024}.

\subsection{Bayesian retrieval}\label{sec:bayesian_retrieval}

We sampled the posterior distribution with nested sampling as implemented in PyMultiNest \citep{buchnerPyMultiNestPythonInterface2016}. We ran each night independently with 800 live points and a constant efficiency of 5\%, following common settings in the literature \citep{zhang12CO132021,regtESOSupJupSurvey2024}.

The free parameters comprised the atmospheric structure (\Teff, \logg, [M/H], $f_{\rm sed}$, $\log K_{zz}$), cloud properties (cloud-base pressure, cloud-base mass fraction, and opacity from Fe and Mg$_2$SiO$_4$ condensates; \citealt{vosPatchyForsteriteClouds2023, morleySonoraSubstellarAtmosphere2024}), planetary radius, radial velocity, \vsini, C/O ratio, and \Cratio. A noise inflation parameter $b$ accounted for underestimated uncertainties or model--data mismatches. It was defined as $\sigma_{\rm scaled}^2=\sigma_{\rm data}^2 + 10^{b}$, ensuring correct error propagation into the posteriors \citep{lineUniformAtmosphericRetrieval2015,molliereEvidenceSiOCloud2025}.

We adopted the likelihood of \citet{landmanPictorisEyesUpgraded2024} (following \citealt{ruffioRadialVelocityMeasurements2019}) and evaluated it separately for each night and nod position, as in \citet{gonzalezpicosESOSupJupSurvey2025}. For a given night and nod,
\begin{align}\label{eq:likelihood}
    \log \mathcal{L} &= - \frac{1}{2} \left[ N \log(2\pi) + \log\det\mathbf{C} + (\mathbf{d} - \mathbf{m})^\top \mathbf{C}^{-1} (\mathbf{d} - \mathbf{m}) \right], \\
    \mathbf{m} &= \mathbf{M}\boldsymbol{\phi},
    \quad
    \boldsymbol{\phi} = \arg\min_{\boldsymbol{\phi}\geq 0} \left\| \mathbf{d} - \mathbf{M}\boldsymbol{\phi} \right\|_2^2,
\end{align}
where $\mathbf{d}$ is the data vector and $\mathbf{C}$ is the covariance built from the scaled uncertainties. For each order--detector spectrum, the design matrix $\mathbf{M}$ contains the planet atmospheric spectrum with telluric transmission ($\mathbf{m}_0$) and $n_{\rm star}$ shifted spline components of the on-axis stellar spectrum \citep{gonzalezpicosESOSupJupSurvey2025}. We solve for $\boldsymbol{\phi}$ at each likelihood evaluation with \texttt{scipy.optimize.nnls} \citep{lawsonSolvingLeastSquares1995}, so stellar contamination and low-frequency systematics are constrained without extra non-linear parameters. The planet amplitude $\phi_0$ gives the wavelength-dependent planet-to-star flux ratio, which we use to track night-to-night variations driven by observing conditions and adaptive optics (AO) performance.

\section{Results}\label{sec:results}
We present the combined spectrum and average best-fit model in \Cref{fig:fig_spectrum_13CO_chips10-11_A}, together with the cross-correlation detection of \thirteenCO. The data quality varies substantially between nights (S/N per pixel between 2 and 9 after error scaling), which we attribute to changing observing conditions. Individual-night constraints on \Teff, [M/H], and C/O are shown in \Cref{fig:fig_bma_teff_metallicity_c_to_o}, and those on \Cratio\ are shown in \Cref{fig:fig_bma_log_12CO_13CO}. Nights with S/N\,$>$\,5 agree within the uncertainties, while lower-S/N nights show systematic shifts. We therefore combined the posterior distributions of the six best nights (S/N\,$>$\,5) with equal weights to obtain the final constraints and incorporate the night-to-night scatter into the uncertainty budget. Alternative weighting schemes, such as Bayesian model averaging \citep{nixonMethodsIncorporatingModel2024} or S/N-weighted combination, assign negligible weight to all but the highest-quality nights. We therefore adopted equal weights and set a threshold of S/N\,$>$\,5 to retain the scatter as a conservative estimate of systematic uncertainties for nights with comparable S/N.

\subsection{Atmospheric parameters}

\subsubsection{Temperature structure}
From the joint posterior, we obtain \Teff$\,=1629^{+30}_{-28}$\,K. This is in line with the spectral type of \betapicb\ and with earlier SED and low-resolution spectroscopic analyses \citep{bonnefoyNearinfraredSpectralEnergy2013, baudinoInterpretingPhotometrySpectroscopy2015, worthenMIRIMRSObservations2024, ravetMultimodalAtmosphericCharacterization2025}. The retrieved pressure--temperature profiles systematically reach the lower end of the grid at \fsed$\,\leq1$ (\Cref{fig:fig_cornerplot_selected_nights}), so the best-fitting P--T structures match models that include thick clouds. For nights with S/N\,$>$\,5, the retrieved \Teff\ values agree well with the adopted value; lower-S/N epochs show larger scatter but remain consistent with it within 1$\sigma$. Our \Teff\ is inferred directly from the P--T profile fit. It is therefore model-dependent and sensitive to the cloud-opacity assumptions built into the Sonora Diamondback grid \citep{morleySonoraSubstellarAtmosphere2024}. Estimates of \Teff\ from a full spectral energy distribution should be more reliable, because they provide an empirical reading of the bolometric flux.

\subsubsection{Chemical composition}
We recover [M/H]$\,=0.20^{+0.16}_{-0.12}$ and C/O$\,=0.52\pm0.03$. Both are compatible with recent work that favours roughly solar to mildly super-solar metallicity and near-solar C/O \citep{ravetMultimodalAtmosphericCharacterization2025}, but they disagree with the metallicity and chemistry inferred from GRAVITY low-resolution spectroscopy \citep{gravitycollaborationPeeringFormationHistory2020}. Our C/O exceeds the value reported from \CRIRES\ by \citet{landmanPictorisEyesUpgraded2024} (C/O$\,=0.41\pm0.04$). Their limited constraints on metallicity, with subsolar [M/H] in all models except the one that adopted the GRAVITY P--T structure, likely propagate into C/O, which cautions against a direct comparison. The C/O ratios from individual nights approach the solar value ($\approx 0.59$; \citealt{asplundChemicalMakeupSun2021}) as the S/N increases (right panel of \Cref{fig:fig_bma_teff_metallicity_c_to_o}). Recent C/O measurements for young brown dwarfs in the $\beta$~Pictoris moving group \citep{liuChemistryIsotopeRatios2026} suggest near-solar compositions, consistent with our findings and with other young brown dwarfs \citep{gonzalezpicosESOSupJupSurvey2024,grasserESOSupJupSurvey2025}.

\subsubsection{Cloud properties}
The spectra show tentative evidence for condensate opacity: the retrieval allows non-zero contributions from Fe and Mg$_2$SiO$_4$ clouds, with posteriors that differ between epochs (\Cref{fig:fig_cornerplot_selected_nights}). A tentative SiO detection in the M band by \citealt{parkerRedMbandStudy2024} suggests that not all silicon is locked in silicates, which would favour Fe as the dominant cloud-forming species. Unlike \citet{landmanPictorisEyesUpgraded2024}, who found cloud parameters to be largely unconstrained and argued for cloud-free solutions because their P--T structures did not cross any of the considered condensation curves, we treated the cloud-base pressure and mass fraction of each species as free parameters. We therefore did not tie condensation strictly to optical constants and the intersection of the P--T profile. Several high-S/N nights favour cloud opacity near the bottom of the photosphere ($P_{\rm base}\approx1$\,bar) with elevated base mass fractions ($\log X_{\rm base}\approx-4$ to $-1$), whereas lower-S/N nights yield weak or absent constraints. A full treatment of cloud composition is left to future work, which would benefit from a broader wavelength range, in particular at bluer wavelengths where the spectrum is more sensitive to cloud properties (e.g. \citealt{zhangELementalAbundancesPlanets2023a}).

\subsubsection{Spin}
The retrieved \vsini\ values are in the range of $18$--$19$\,\kms\ and agree broadly with recent high-resolution estimates ($19.9 \pm 1.0$\,\kms; \citealt{landmanPictorisEyesUpgraded2024}), but they are systematically lower than some earlier values \citep[e.g.][]{snellenFastSpinYoung2014}. We attribute the difference mainly to the rotational broadening implementation, which integrates the visible hemisphere and includes angle-dependent intensities rather than a single-temperature slab \citep{mollierePetitRADTRANSPythonRadiative2019}. A convolution kernel with a limb-darkening correction may overestimate \vsini. Applying the rotational broadening to intensity spectra at different incident angles gives a \vsini\ value that naturally includes limb darkening. Small differences in \vsini\ between nights may reflect surface heterogeneity, which can alter the line profiles, but we do not explore this further in the present work.

\subsection{Carbon isotope ratio}

From the combined posterior, we measure \Cratio$\,=58^{+18}_{-15}$ in the atmosphere of \betapicb\ (\Cref{fig:fig_bma_log_12CO_13CO}). This is consistent with the present-day local ISM (\Cratio$\,=\,68\pm14$; \citealt{milam12C13CIsotope2005}) and below the solar photospheric value (\Cratio$_{\odot}\,=\,91.4\pm1.3$; \citealt{lyonsLightCarbonIsotope2018}). An independent GRAVITY+ analysis (\citealt{vonStauffenbergBetaPic2026}; companion paper) finds a somewhat higher ratio, around the solar value and compatible with our measurement within 1$\sigma$. We emphasise the importance of multi-epoch validation of \Cratio\ to ensure that the measurement is not biased by low-S/N data or model--data mismatches. Measurements from individual nights show discrepancies up to 3$\sigma$ (see \Cref{fig:fig_bma_log_12CO_13CO}), which can lead to incorrect conclusions about the planet's carbon isotopic composition if assessed in isolation. This is the case for two of our six nights with S/N\,$>$\,5, which may be affected by systematic errors in the retrieval or data reduction (see \Cref{fig:fig_cornerplot_selected_nights}). The remaining four higher-S/N epochs show good agreement within 1$\sigma$. By combining the posteriors from several epochs of comparable quality, we obtain a more robust measurement, with uncertainties that reflect the night-to-night scatter. Our final adopted value is compatible with all individual night measurements within 1$\sigma$.

\citet{ravetMultimodalAtmosphericCharacterization2025} tentatively inferred much lower \Cratio\ for \betapicb\ (down to $\approx 11$) from a cross-correlation detection of \thirteenCO\ in medium-resolution GRAVITY data ($\mathcal{R}\sim4000$). Such values would imply strong \thirteenC\ enrichment and are neither found in our retrieval nor supported by the GRAVITY+ study \citep{vonStauffenbergBetaPic2026}. Low-S/N detections can bias isotopic ratios towards low \Cratio, as \citet{zhangESOSupJupSurvey2024} illustrated for YSES~1~b by revising a SINFONI-based estimate of 31 to 88 with higher-S/N, higher-resolution \CRIRES\ data. We therefore stress the need to confirm \Cratio\ at adequate resolving power and S/N.

An ISM-like \Cratio\ indicates that the carbon isotopic composition was not substantially altered during planet formation and evolution. At a separation of approximately 10\,au from its A-type host star \citep{lacourMassPictorisPictoris2021}, \betapicb\ lies interior to the CO snow line \citep{obergEFFECTSSNOWLINESPLANETARY2011}. If the planet formed near its current location, the available carbon budget was predominantly in the gas phase and was therefore probably not subject to low-temperature fractionation processes that can enrich \thirteenC\ in ices \citep{woodsCarbonIsotopeFractionation2009}. 

\begin{figure}[ht]
    \centering
    \includegraphics[width=\columnwidth]{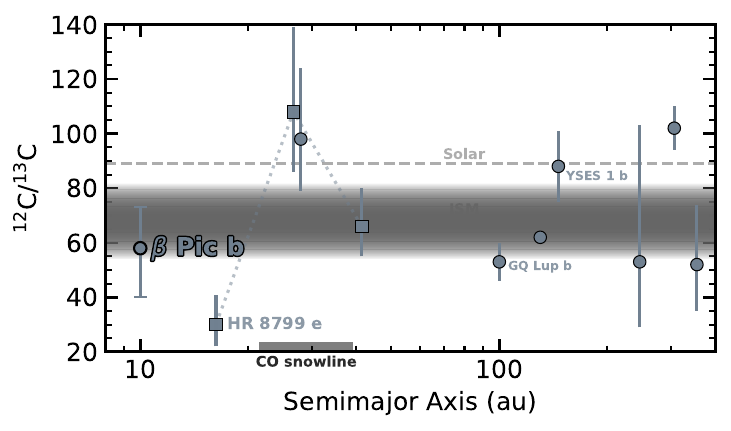}
    \caption{Carbon isotope ratio as a function of the semi-major axis for young, directly-imaged companions. Measurements in the HR 8799 system for planets e, b, and c \citep{ruffioJupiterlikeUniformMetal2026} are shown with squares and connected with a dotted line. The other measurements are for HD 984 b \citep{costesFreshViewHot2024}, GQ Lup b \citep{gonzalezpicosESOSupJupSurvey2025}, VHS 1256 b \citep{gandhiJWSTMeasurements13C2023}, YSES 1 b \citep{zhangESOSupJupSurvey2024}, AB Pic b \citep{gandhiESOSupJupSurvey2025}, and DH Tau b and HIP 79098 b \citep{xuanAreThesePlanets2024}. The present-day local ISM (\Cratio$\,=\,68\pm14$; \citealt{milam12C13CIsotope2005}) and solar photospheric value (\Cratio$_{\odot}\,=\,91.4\pm1.3$; \citealt{lyonsLightCarbonIsotope2018}) are overlaid. The horizontal line spans the typical CO snow-line location for a solar-type host star (22--36 au). For stars hotter than the Sun, the CO snow line is expected at distances greater than 30 au, so \betapicb\ is located interior to the CO snow line.}
    \label{fig:fig_isotope_ratios_semimajor_axis}
\end{figure}

Measurements of carbon isotope ratios in young giant planets so far span \Cratio$\,\approx 30$--120 (see \Cref{fig:fig_isotope_ratios_semimajor_axis}). The literature shows no clear dependence of \Cratio\ on atmospheric properties or orbital distance. Additional measurements with improved precision are needed to assess whether carbon isotope ratios are useful formation tracers. The current sample is small, with around a dozen measurements (see \Cref{fig:fig_isotope_ratios_semimajor_axis}), and the uncertainties are often too large to draw firm conclusions. We expect this sample to grow substantially with next-generation facilities such as the Extremely Large Telescope (ELT)/METIS \citep{brandlMETISMidinfraredELT2021}, which will access the fundamental CO band around 4.6\,\micron\ at high resolution and with much higher S/N.

Future constraints on stellar or disc \Cratio\ would help contextualise our atmospheric measurement. High-resolution spectroscopy of the A-type host remains impractical, but (sub)millimetre interferometry can in principle constrain carbon isotopes in circumstellar gas \citep{cataldiALMAResolvesEmission2018, yoshidaNewMethodDirect2022}. For $\beta$\,Pictoris, the debris disc is collisionally replenished by exocometary material, so millimetre-line \Cratio\ measurements would trace reprocessed solids and released volatiles rather than a pristine nebular reservoir. Isotopic measurements of objects in the same association may therefore provide a useful baseline: a young brown dwarf in the $\beta$~Pictoris moving group, 2MASSI J0443+0002, has a reported value of \Cratio$\,=70\pm5$ \citep{liuChemistryIsotopeRatios2026}, comparable to the ISM value and consistent with our result within the uncertainties.

\section{Conclusion}
We presented eleven epochs of K-band \CRIRES\ spectroscopy of \betapicb\ at $\mathcal{R}\approx100{,}000$, reduced with \texttt{excalibuhr} and extracted with a cross-correlation-weighted scheme suited to faint companions on a starlight-dominated background. A forward model combining the planet, scattered starlight, and telluric absorption, fitted jointly to the high-resolution spectrum and broadband photometry, yields a \thirteenCO\ detection and \Cratio$\,=58^{+18}_{-15}$, consistent with the local ISM and below the solar photospheric value. The retrieval simultaneously constrains the thermal and chemical structure (\Teff$\,=1629^{+30}_{-28}$\,K, [M/H]$\,=0.20^{+0.16}_{-0.12}$, C/O$\,=0.52\pm0.03$, and tentative condensate opacity). These results broadly align with recent work while differing from several earlier \CRIRES\ and GRAVITY-based inferences. The added photometry eases metallicity--cloud degeneracies inherent to high-resolution spectra alone and places the metallicity of \betapicb\ in the solar to mildly super-solar range. Observations of younger, gas-rich protoplanetary discs may provide useful context for linking atmospheric \Cratio\ to formation chemistry when stellar constraints are unavailable. Extending high-resolution isotopic measurements to additional companions in the same system, notably the inner super-Jupiter \betapicc, will eventually test whether in situ \Cratio\ tracks orbital distance together with the population-level benchmarks for young, directly-imaged planets.

\begin{acknowledgements}
We thank the anonymous referee for their helpful comments and constructive suggestions, which significantly improved the manuscript.
D.G.P. and I.S. acknowledge NWO grant OCENW.M.21.010. Support for this work was provided by the NL-NWO Spinoza (SPI.2022.004). Based on observations collected at the European Organisation for Astronomical Research in the Southern Hemisphere under ESO programme(s) 114.27DX.001. This work used the Dutch national e-infrastructure with the support of the SURF Cooperative using grant no. EINF-4556. JLB acknowledges funding from the European Research Council (ERC) under the European Union’s Horizon 2020 research and innovation programme under grant agreement No 805445.\newline
Software: NumPy \citep{harrisArrayProgrammingNumPy2020}, SciPy \citep{virtanenSciPyFundamentalAlgorithms2020}, Matplotlib \citep{hunterMatplotlib2DGraphics2007}, petitRADTRANS \citep{mollierePetitRADTRANSPythonRadiative2019}, species \citep{stolkerMIRACLESAtmosphericCharacterization2020}, fastchem \citep{kitzmannFastchemCondEquilibrium2023}, PyMultiNest \citep{buchnerPyMultiNestPythonInterface2016}, Astropy \citep{collaborationAstropyProjectSustaining2022}, corner \citep{foreman-mackeyCornerPyScatterplot2016}, ExoMol \citep{tennyson2024ReleaseExoMol2024}, HITEMP \citep{rothmanHITEMPHightemperatureMolecular2010}, pyROX \citep{regtPyROXRapidOpacity2025}.
\end{acknowledgements}
\bibliography{autolibrary}
\onecolumn

\begin{appendix}

\section{Posterior distributions}

\begin{figure}[ht]
    \centering
    \includegraphics[width=\textwidth]{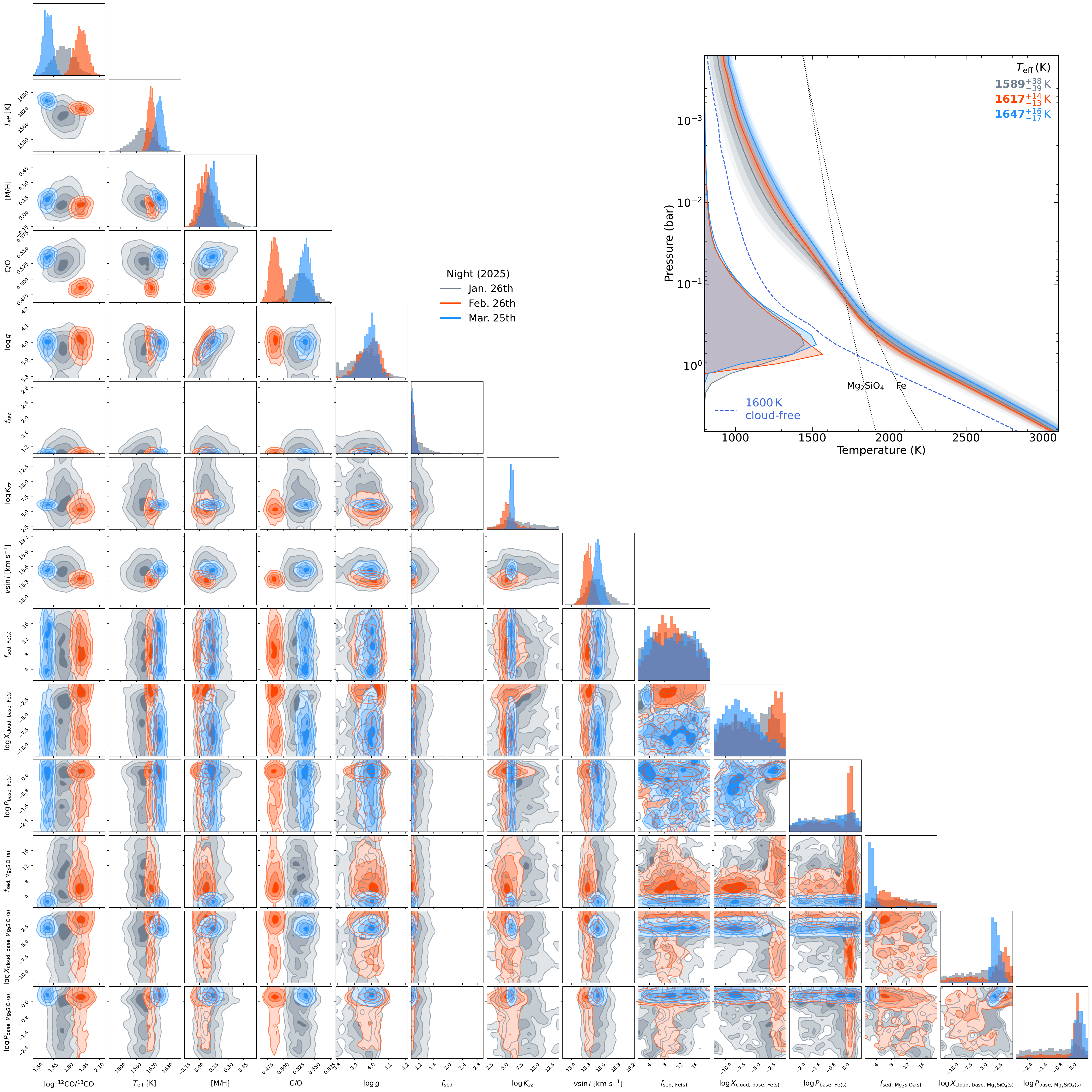}
    \caption{Posterior distributions for three representative nights: January 26 (S/N$\approx$3), February 26 (S/N$\approx$7), and March 25 (S/N$\approx$8), 2025. The two higher-S/N nights were deliberately chosen from the S/N\,$>$\,5 subset to illustrate that even high-S/N single-epoch retrievals can show significant discrepancies in some atmospheric parameters, underscoring the precision limitations of single-epoch analysis with moderate-S/N data. Additional scatter may arise from degeneracies between metallicity, surface gravity, and cloud properties. The sedimentation parameter $f_{\rm sed}$ used for interpolating the Sonora Diamondback P--T profiles consistently reaches the grid lower bound ($f_{\rm sed}\leq1$), corresponding to thick clouds. Cloud-opacity sedimentation is fitted separately for each cloud species, but remains weakly constrained and varies between nights. The upper panel shows the 1, 2, and 3-$\sigma$ temperature envelopes for each night, with Fe and Mg$_2$SiO$_4$ condensation curves overlaid (dotted). Shaded regions show the integrated emission contribution functions, indicating the vertical extent of the photosphere.}
    \label{fig:fig_cornerplot_selected_nights}
\end{figure}

\end{appendix}

\end{document}